\documentclass[prl,10pt,twocolumn,superscriptaddress,floatfix,showpacs]{revtex4-2}

\usepackage[utf8]{inputenc}  
\usepackage[T1]{fontenc}   
\usepackage[british]{babel} 
\usepackage[scaled=0.86]{berasans} 
\usepackage[colorlinks=true, allcolors=blue, urlcolor=blue]{hyperref} 
\usepackage{graphicx} 
\usepackage[babel]{microtype} 
\usepackage{amsmath,amssymb,amsthm,bm,amsfonts,mathrsfs,bbm} 

\usepackage{xspace} 
\usepackage{pgfplots}
\usepackage{xcolor,colortbl}
\usepackage{array}
\usepackage{bigstrut}
\usepackage{comment}

\newcommand{\bpm}{\begin{pmatrix}}
\newcommand{\epm}{\end{pmatrix}}
\newcommand{\tr}{\mathrm{tr}}
\newcommand{\ket}[1]{| #1 \rangle}
\newcommand{\bra}[1]{\langle #1|}
\newcommand{\ip}[2]{\langle #1|#2 \rangle}
\newcommand{\ketbra}[1]{\ket{#1}\!\bra{#1}}
\newcommand{\be}{\begin{equation}}
\newcommand{\ee}{\end{equation}}
\newcommand{\bea}{\begin{eqnarray}}
\newcommand{\eea}{\end{eqnarray}}
\newcommand{\bes}{\begin{equation*}}
\newcommand{\ees}{\end{equation*}}
\newcommand{\bean}{\begin{eqnarray*}}
\newcommand{\eean}{\end{eqnarray*}}

\newcommand{\x}{\mathrm{x}}
\def\y{\mathrm{y}}

\def\I{\mathbbm{I}}

\newcommand{\id}{\mathbb{I}}

\newtheorem*{thm*}{Theorem}

\newtheorem*{lem*}{Lemma}

\newtheorem*{lipschitzLem*}{Lemma \ref{lipschitz}}
\newtheorem*{lipschitzCubeLem*}{Lemma \ref{lipschitzCube}}
\newtheorem*{pgmNearlyOptimalThm*}{Theorem \ref{pgmNearlyOptimal}}

\begin{document}
\title{Semi-Device-Independent Certification for Nonlocality without Entanglement}

\author{Hanwool Lee }
\email{hanwool.h.lee@jyu.fi}
\affiliation{Faculty of Information Technology, University of Jyväskylä, Finland}

\author{Joonwoo Bae}
\email{joonwoo.bae@kaist.ac.kr}
\affiliation{School of Electrical Engineering, Korea Advanced Institute of Science and Technology (KAIST), 291 Daehak-ro, Yuseong-gu, Daejeon 34141, Republic of Korea }

\begin{abstract}


In this work, we present the framework for demonstrating and certifying the distinction between measurements in an entangled basis, called global measurements, and local operations and classical communication (LOCC), known as nonlocality without entanglement (NLWE). To be precise, we show NLWE via a maximum-confidence measurement in terms of a guess {\it per} detection event, called a {\it confidence}, a fine-grained guessing probability that encompasses both minimum-error and unambiguous state-discrimination strategies. We show that NLWE for unknown measurements can be certified, given the measurement-outcome rates, by bounding the confidence using LOCC; the certification is semi-device-independent in that state preparation is trusted. We illustrate the demonstration and certification of NLWE for antiparallel qubit states. Our results make it feasible to experimentally realize NLWE using measurement devices with imperfections, such as non-unit detection efficiency, since maximum-confidence measurements rely only on detected events. 


\end{abstract} 

\maketitle

Quantum information processing in its most fundamental setting contains {\it state preparation} and {\it measurements} such that it may go beyond the capabilities of the classical counterpart. The distinction between local operations and classical communication (LOCC) and global operations in {\it state preparation} characterizes entangled states, which have been identified as a key resource for quantum advantages \cite{RevModPhys.81.865}. Nonlocality without entanglement (NLWE) signifies a gap existing between LOCC and measurements in an entangled basis, denoted by GLOBAL throughout \cite{bennett1999quantum}. Namely, there are ensembles of states with no entanglement for which GLOBAL outperforms LOCC. NLWE is also closely related to entanglement witnesses \cite{Ha_2023}, a standard method for verifying entanglement \cite{TERHAL2002313, PhysRevA.62.052310,GUHNE20091, Chruscinski_2014, sgbs-7228}. Recently, the certification of NLWE has been presented in a device-independent manner \cite{PhysRevA.107.062220}. 

Quantum state discrimination, one of the most fundamental tasks connected to various applications \cite{bergou2010discrimination, barnett2009quantum, bae2015quantum}, has been used to demonstrate NLWE. The distinction between GLOBAL and LOCC is shown in terms of the guessing probability in minimum-error discrimination for an ensemble of states, such as double-trine states \cite{peres1991optimal, chitambar2013revisiting} and domino states \cite{bennett1999quantum}. Ensembles showcasing NLWE have been generalized \cite{ Ha:2021aa}. Furthermore, unambiguous discrimination has been used to show a gap between GLOBAL and separable measurements, denoted by SEP, such that the rate of inconclusive outcomes can be lower with GLOBAL than SEP \cite{chitambar2013local}.  

The strategies mentioned above for state discrimination contain stringent constraints. Minimum-error discrimination does not exploit an inconclusive outcome; all are conclusive and used to make a guess \cite{helstrom1967detection}. Unambiguous discrimination asks for zero-error state identification by conclusive outcomes \cite{dieks1988overlap}, see also reviews \cite{barnett2009quantum, bergou2010discrimination, bae2015quantum}. However, a detector, particularly in a realistic setting, generally produces conclusive and inconclusive results, and both can be used to demonstrate quantum advantages \cite{PhysRevX.8.011015, flatt2022contextual, Flatt_2026, 196j-lmzl}. In addition, a detector with imperfections will not satisfy the zero-probability constraint for either a conclusive or an inconclusive outcome. Then, imperfections may lead to loopholes, such as in Bell nonlocality \cite{PhysRevA.47.R747, PhysRevLett.98.220403}.

The present work aims to establish a framework for both {\it demonstrating} and {\it certifying} NLWE in a practical scenario in which a measurement device yields both conclusive and inconclusive outcomes. That is, both conclusive and inconclusive outcomes can be used to demonstrate and certify NLWE. To this end, we consider the maximum-confidence measurement (MCM) strategy, which generalizes minimum-error and unambiguous discrimination. In the certification scenario, we show that NLWE can be certified for unknown measurements; hence, it corresponds to a semi-device-independent (sDI) scenario. 
 
 \begin{figure} \label{fig:scheme}
    \centering
    \includegraphics[width=0.9 \linewidth]{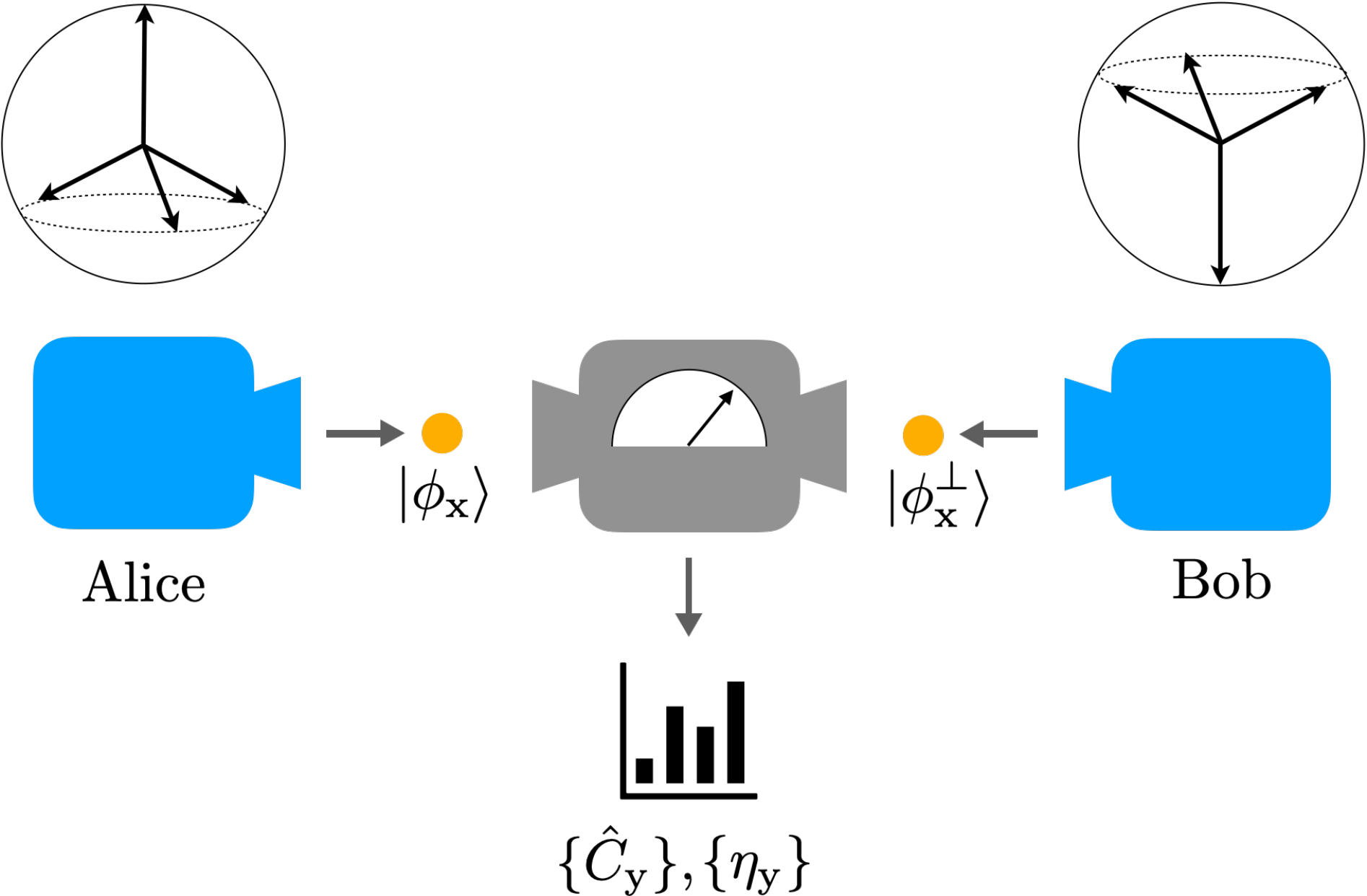}
    \caption{ The scenario of comparing the capabilities of measurements in GLOBAL and SEP is shown. Alice and Bob prepare unentangled states by LOCC, such as antiparallel states in Eq. (\ref{eq:antitetraens}). NLWE can be demonstrated and certified if an optimal measurement is GLOBAL but not SEP. }     \label{fig:placeholder}
\end{figure}

To demonstrate the framework, we consider an ensemble of antiparallel qubit states. On the one hand, we demonstrate NLWE by showing that GLOBAL gives a higher confidence on a guessing task than SEP or LOCC. Since an MCM finds confidence in individual detection events, the distinction between GLOBAL and SEP is shown in a fine-grained guessing task that encompasses the strategies of minimum-error and unambiguous state discrimination. On the other hand, we show the certification of an NLWE after collecting detection events. We reiterate that the certification works while a measurement is not yet verified. The framework for certifying NLWE applies to measurement devices that contain imperfections, including undetected outcomes due to photon loss or low detection efficiency.

Let us first summarize MCMs \cite{PhysRevLett.96.070401}. While a quantum measurement is in a prediction form, allowing one to compute probabilities of outcomes given a state, an MCM, conversely, is retrodictive in that an outcome makes a guess about preparations \cite{Barnett:00, sym13040586}. Let $S =\{ q_{\x}, \rho_{\x}\}$ denote an ensemble i.e., a state $\rho_{\x}$ with {\it a priori} probability $q_{\x}$. The conditional probability in the following is called a {\it maximum confidence} (MC), 
\cite{PhysRevLett.96.070401}, 
\bea \label{eq:globalmc}
C_{\x,\max}:=\max_M p_{P|M}(\x | \x)=  \max_{M_\x \geq 0 }\frac{q_\x \tr[\rho_\x M_\x]}{\tr[ \rho M_\x]}, 
\eea
where $P$ and $M$ denote preparation and measurement, respectively. 

For instance, we have $C_{\x,\max}=1$ if an outcome $\x$ implies the preparation of a state $\rho_{x}$ with certainty; this case also corresponds to unambiguous discrimination. Note that the guessing probability in minimum-error discrimination can be reproduced as an average MC, i.e., $p_{\mathrm{guess}}= \sum_{\x}  \tr[\rho M_{\x}] C_{\x}$ when all outcomes are conclusive \cite{barnett2009quantum}. Hence, an MC above can be interpreted as a fine-grained guessing probability unifying both well-known strategies of state discrimination \cite{barnett2009quantum}.

The optimization problem in Eq. (\ref{eq:globalmc}) can be in a linear form \cite{barnett2009quantum} or formulated as a semidefinite program (SDP) \cite{lee2022maximum}. It turns out that $C_{\x, \max} = \|\sqrt{\rho}^{-1} q_\x \rho_\x \sqrt{\rho}^{-1} \|_\infty$ with an operator norm $\|\cdot \|_{\infty}$. Therefore, an MCM can be generally implemented by rank-one POVM elements. 

Let $C_{\x, \max}^{({G})}$ denote an MC with measurements in GLOBAL, and $C_{\x, \max}^{( S)}$ with measurements in SEP. Note that SEP can be generally implemented via stochastic LOCC \cite{chitambar2014everything}. For a separable POVM element $M_\x$, there exists $\epsilon>0$ such that $\epsilon M_\x$ can be realized by LOCC. Since an MC remains identical under scaling a POVM element \cite{flatt2022contextual}, we have $C_{\x, \max}^{(\mathrm{LOCC})} = C_{\x, \max}^{({S})}$; an MCM does not find a gap between LOCC and SEP. 

Then, MCMs in Eq. (\ref{eq:globalmc}) can be obtained by analyzing the optimality conditions of the SDP. From the approach of the complementarity problem \cite{lee2022maximum}, see also \cite{Bae_2013} for the minimum-error case, the optimization problem can be reformulated to find optimal parameters, a POVM element $M_{\x}$ and an ensemble of complementary states $\{r_{\x}, \sigma_{\x}\}$, where $r_{\x}\geq 0$, that satisfy the equality conditions
\bea \label{eq:optimality}
\tr[\sigma_\x M_\x ]=0 ~\mathrm{and}~ C_{\x, \max}^{(G)}\rho-q_\x \rho_\x=r_\x \sigma_\x, 
\eea
in which $C_{\x,\max}^{(G)}$ is obtained from the equality condition. A POVM element may be constrained to be SEP and finds an MC with SEP, $C_{\x,\max}^{(S)}$.

In Eq. (\ref{eq:optimality}), once the latter condition finds complementary states, a POVM element can be constructed from the former one, called the complementarity slackness. In this case, SEP is optimal if there is a product vector $|e\rangle|f\rangle$, i.e., $M_{\x}^{(S)} =|e\rangle\langle e| \otimes |f\rangle \langle f|  $, such that $\tr[\sigma_{\x} M_{\x}^{(S)}]=0$. In other words, if $\tr[\sigma_{\x}M_{\x}^{(S)}] >0$ for all $M_{\x}^{(S)}\in\mathrm{SEP}$, an optimal POVM element is not in SEP but in GLOBAL; hence, it follows $C_{\x,\max}^{(G)} > C_{\x,\max}^{(S)}$. Let us summarize the necessary and sufficient condition for finding a gap between GLOBAL and SEP. \\

\textbf{Proposition 1.} In the complementary problem to finding an MCM for an ensemble $S$, consisting of complementary states, denoted by $\{\sigma_{\x} \}$, it holds $C_{\x, \max}^{(\mathrm{G} )}=C_{\x, \max}^{( \mathrm{S} )}$ for an outcome $\x$ if and only if there exists a product vector $|e\rangle|f\rangle$ such that $\sigma_\x |e\rangle|f\rangle =0$. If no product state satisfies the condition, the ensemble $S$ can exhibit the distinction between GLOBAL and SEP in terms of an MC, i.e., $C_{\x, \max}^{(\mathrm{G})} > C_{\x, \max}^{(\mathrm{S})}$. \\

In the following, we apply the general results above to an ensemble of two-qubit antiparallel states, 
\bea \label{eq:antitetraens}
S_{\perp} = \{ \ket{\Psi_\x}\}_{\x=1}^4~\mathrm{where}~ \ket{\Psi_\x}=\ket{\phi_\x}\otimes \ket{\phi_\x^\perp},
\eea
where an ensemble $\{|\phi_{\x} \rangle\}$ corresponds to symmetric, informationally complete states
\bea
&& \ket{\phi_1}=\ket{0}, ~~\ket{\phi_2}=\frac{1}{\sqrt{3}}\ket{0}+\sqrt{\frac{2}{3}}\ket{1}, \nonumber \\
&&\mathrm{and}~~ \ket{\phi_{3,4}}=\frac{1}{\sqrt{3}}\ket{0}+e^{\pm \frac{2 \pi i}{3}}\sqrt{\frac{2}{3}}\ket{1}. \nonumber 
\eea 
Note that the ensemble has been used to demonstrate NLWE in state estimation \cite{gisin1999spin} in that optimal state estimation is GLOBAL on antiparallel states, whereas it is SEP on parallel states. Antiparallel and parallel ensembles are related by a universal NOT gate \cite{PhysRevA.60.R2626}, which is antiunitary and thus nonphysical. Then, state estimation is more efficient for antiparallel states, for which an optimal measurement is GLOBAL, than for parallel ones, for which it is SEP.

For the ensemble above, we compute the MC and have $C_{\x,\max}^{(G)}=1$ with a POVM element $M_\x=a_\x \ketbra{\varphi_\x}$ where
\bea\label{eq:glomcmanti}
\ket{\varphi_\x}=\frac{1}{\sqrt{5}} (2\ket{\phi_\x}\otimes\ket{\phi_\x^\perp}+\ket{\phi^\perp_\x}\otimes\ket{\phi_\x}).
\eea
for some $a_{\x} \in (0,1]$. Note that unambiguous discrimination is realized since $\ip{\Psi_\y}{\varphi_\x}=0$ for $\x \neq \y$. Thus, it is clear that $C_{\x,\max}^{(S)}<1$ with SEP. 


A separable measurement for the MC is also rank-one. With an ansatz, $M_\x = b_{\x} \ketbra{e_\x} \otimes \ketbra{f_\x}$ for some $b_{\x}\in(0,1]$, we compute 
\bea\label{eq:sepmctetra}
C_{\x, \max}^{({S})}
=\max_{\ket{e_\x},\ket{f_\x}} \frac{|\ip{\phi_\x}{e_\x}\ip{\phi_\x^\perp}{f_\x}|^2}{\sum_\y |\ip{\phi_\y}{e_\x}\ip{\phi_\y^\perp}{f_\x}|^2}.
\eea
Since $|\ip{\phi_\y^\perp}{f_\x}|=|\ip{\phi_\y}{f^\perp_\x}|$, the optimization is equivalent to finding the separable MCM for the parallel states. 
Therefore, we find that the optimal POVM element in SEP for MCM on the ensemble of the antiparallel states is given by 
\bea 
M_\x^{(\mathrm{S})}=b_\x \ketbra{\phi_\x}\otimes \ketbra{\phi_\x^\perp},~~\mathrm{for~some~}b_{\x}>0  \label{eq:sepmcm}
\eea
and have $C_{\x, \max}^{({S})} = 3/4$. Thus, we have demonstrated NLWE for antiparallel states in terms of MCMs. 

We remark that an MCM for an ensemble of the parallel states $S_{\|} = \{ |\phi_{\x}\rangle^{\otimes 2} \}_{\x=1}^4$ can be achieved by measurements in SEP. The ensemble can be described on a symmetric subspace that can be implemented with SEP; a POVM element in SEP suffices to find an MC, which is $3/4$ for all $\x$. Hence, an ensemble of parallel states cannot be used to distinguish between GLOBAL and SEP. The conclusion aligns with the result that the state estimation is more efficient when a measurement is GLOBAL, for which states are prepared to be antiparallel rather than parallel
\cite{gisin1999spin}.

{\it The sDI certification of NLWE.} We now show that NLWE can be certified using outcomes of uncharacterized measurement devices. The scenario is similar to that considered in Ref. \cite{flatt2022contextual}, where a high rate of detection events is used to rule out certain properties of a measurement. For instance, unambiguous discrimination shows inconclusive outcomes at least with the minimum probability known as the Ivanovic-Dieks-Peres (IDP) limit \cite{ivanovic1987differentiate, dieks1988overlap, peres1988differentiate}. A measurement by which a rate of inconclusive outcomes is too small excludes the possibility of unambiguous discrimination. 

To be precise, the framework contains three parameters. Firstly, an ensemble of states should be specified $S=\{q_{\x},\rho_{\x} \}$. Secondly, one collects measurement outcomes and finds the outcome rates, denoted by $\{\eta_{\x}\}$. For an ensemble $\rho=\sum_{\x}q_{\x}\rho_{\x}$, an outcome rate should be described with a POVM element $M_{\x}$,
\bea
\mathrm{outcome~rate:~~}\eta_{\x} = \tr[M_{\x} \rho] \label{eq:outr}
\eea
for each $\x$. Note also that a probability of an outcome $\x$ given a state $\rho_{\x}$, denoted by $p_{M|P} (\x| \x)$, can be found from state preparations and measurement outcomes, $p_{M|P} (\x| \x) = \tr[M_{\x} \rho_{\x}]$. Thirdly, from these, a confidence can be computed, called certifiable confidence as follows,  
\bea
\mathrm{certifiable~confidence:~~} \hat{C}_\x = \frac{q_{\x} }{\eta_{\x}} p_{M|P} (\x | \x ). \label{eq:cercon}
\eea
With two parameters $\eta_\x$ and $\hat{C}_{\x}$ achieved experimentally, the task to certify NLWE is to find whether a certifiable confidence can be achieved by measurements in SEP.

The SDP for the certifiable MC works when an outcome rate is given,
\bea
\mathrm{certifiable ~ MC}: && \hat{C}_{\x,\max} =   \mathrm{max} ~~ \frac{q_\x }{ \eta_{\x} } \tr[\rho_\x M_\x] \label{eq:sdpconc} \\
&& \mathrm{subject~to}~~\tr[\rho M_\x]=\eta_\x \nonumber \\
&& ~~~~~~~~~~~~~~~ 0 \leq M_\x \leq \id. 
\label{eq:sep}
\eea
Compared to Eq. (\ref{eq:globalmc}), the optimization above finds the achievable MC given outcome statistics. With the constraint that a measurement is SEP or GLOBAL, we write by $\hat{C}_{\x,\max}^{(S)}$ and $\hat{C}_{\x,\max }^{(G)}$, respectively. Then, we conclude NLWE for an unknown measurement if a certifiable confidence is found in the range,
\bea
\hat{C}_{\x,\max}^{(\mathrm{G})} \geq \hat{C}_{\x} > \hat{C}_{\x,\max}^{(\mathrm{S})}, \label{eq:certinq}
\eea
i.e., a certifiable confidence cannot be achieved by SEP. 

To compute a certifiable MC, we exploit the complementarity approach. The dual to Eq.(\ref{eq:sdpconc}) can be derived, 
\bea
\hat{C}_{\x,\max} &=& \mathrm{min~} \tr[N] + \lambda \eta_{\x} \nonumber\\
&& \mathrm{subject~to}~~ N \geq 0, N+\lambda \rho \geq \rho_{\x}. \nonumber
\eea
where $\tr[\rho M_\x]=\eta_\x$. Note that the strong duality holds. The complementarity problem collects the optimality conditions with parameters $P, N \geq 0$ as follows, together with primal and dual constraints, 
\bea 
&& P-N =\lambda \rho - \rho_{\x} \label{eq:ls} \\
&& M_{\x} P =0~~\mathrm{and}~~(\id - M_{\x}) N =0, \label{eq:cs} 
\eea
where the first condition is called the Lagrangian stability and the second one the complementary slackness. The derivation is detailed in the Supplemental Material. With the parameters above, the certifiable MC with GLOBAL can be found, 
\bea
\hat{C}_{\x,\max}^{(G)} =q_{\x} \left( \lambda + \frac{\tr[N]}{\eta_{\x}} \right) \label{eq:cmcg}
\eea
where $\lambda$ and $N$ are dual parameters.



Let us consider product states, $ S = \{ q_{\x}, |\Psi_{\x}\rangle \langle \Psi_{\x} | \}$ with $|\Psi_{\x}\rangle = |\psi_{\x} \rangle_1 |\psi_{\x}^{'}\rangle_2$; an ensemble of antiparallel states in Eq. (\ref{eq:antitetraens}) is an instance. Suppose that a POVM element is of full rank, meaning a high outcome rate. Then, we directly solve the maximization in Eq. (\ref{eq:sdpconc}), and find an optimal POVM element as follows,
\bea
M_{\x} = \alpha_{\x} |\Psi_{\x}\rangle \langle \Psi_{\x} | + (1-\alpha_{\x} ) \id \label{eq:optM}
\eea
which is in SEP where $\alpha_{\x}>0$. The proof is detailed in the Supplemental Material. Given an outcome rate $\eta_{\x}$, it holds that $\eta_{\x} > \langle \Psi_{\x} | \rho | \Psi_{\x}\rangle$. Then, optimal parameters are obtained as follows. From Eqs. (\ref{eq:ls}) and (\ref{eq:cs}), we have $P=0$, $N=\rho_{\x}$, and $\lambda=0$.  Note that these optimal parameters are also valid when $\alpha_{\x}=1$ in Eq. (\ref{eq:optM}); this corresponds to the case $\eta_{\x} = \langle \Psi_{\x} | \rho | \Psi_{\x}\rangle$. Therefore, optimal POVMs belong to SEP and cannot demonstrate NLWE. We summarize a necessary condition to certify NLWE as follows. 
\bigskip

\textbf{Proposition 2.}
{For an ensemble of product states $\{q_{\x}, |\Psi_{\x}\rangle \langle \Psi_{\x}| \}$ where $|\Psi_{\x}\rangle = |\psi_{\x}\rangle_1 |\psi_{\x}'\rangle_2$, NLWE in terms of MCMs cannot be demonstrated for high values of an outcome rate such that $\eta_\x \geq \bra{\Psi_\x} \rho \ket{\Psi_\x}$.
}
\bigskip

\begin{figure}
    \centering
    \includegraphics[width=0.95\linewidth]{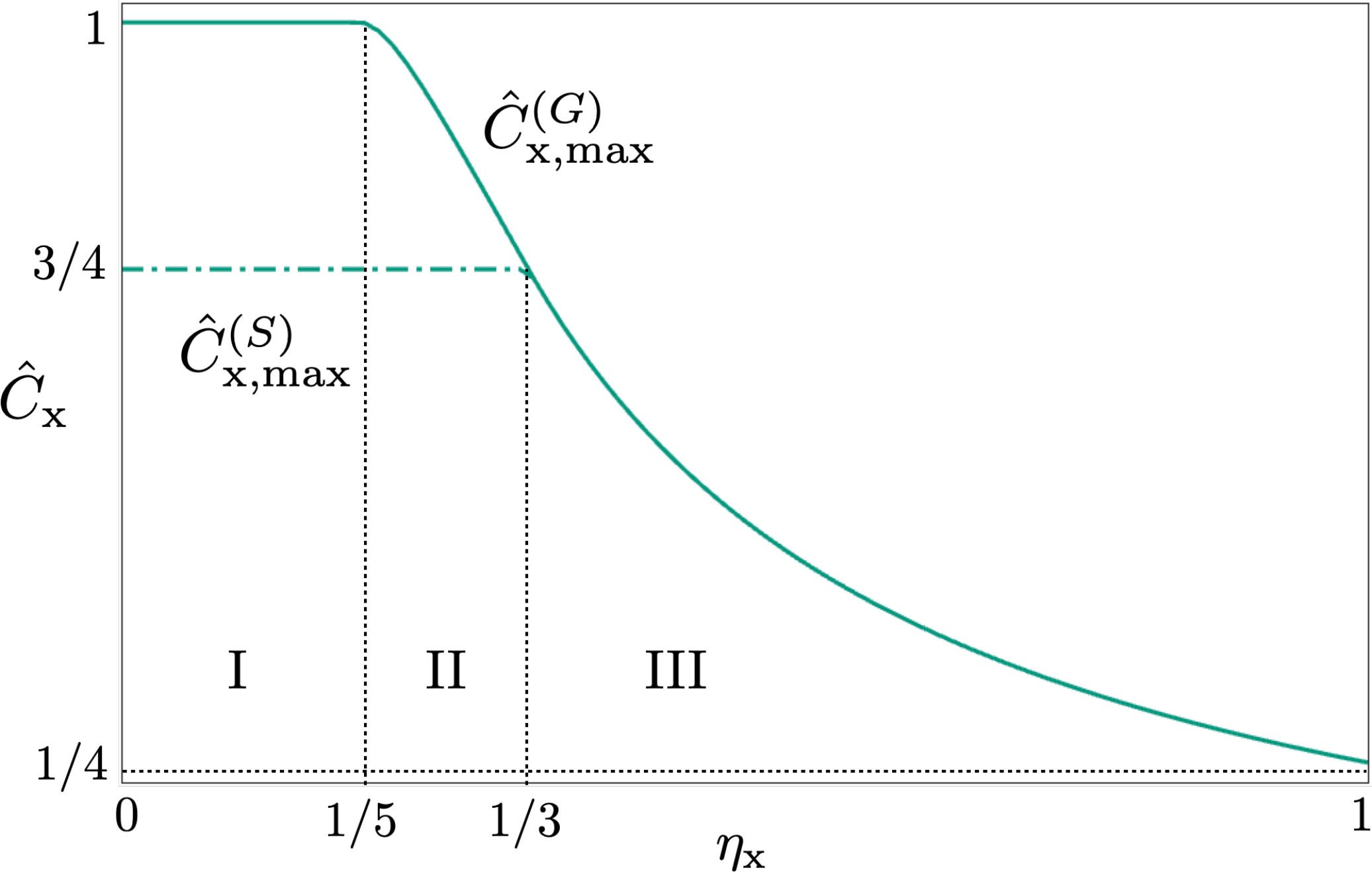}
    \caption{ Given an outcome rate $\eta_{\x}$, the certifiable MC for antiparallel states in Eq. \eqref{eq:antitetraens} is shown. In a noiseless state preparation, the certifiable MC with SEP is $3/4$. For an outcome rate $\eta_{\x}<1/3$, NLWE can be certified if $\hat{C}_{\x}>3/4$ (region I and II). For an outcome rate $\eta_\x \leq 1/5$, the gap $\Delta_{\x}$ is maximal (region I). When outcomes are so frequent $\eta_{\x}\geq 1/3$, NLWE cannot be certified (region III).
    }
    \label{fig:conc2}
\end{figure}

Therefore, to certify NLWE, it is necessary that the outcome rate should be limited such that $\eta_{\x} < \langle \Psi_{\x} | \rho | \Psi_{\x}\rangle$. The condition applies to antiparallel states in Eq. (\ref{eq:antitetraens}) and we have $\eta_\x < 1/3$. 

Let us analyze the optimality condition in Eq. (\ref{eq:ls}) as follows,
\bea
\sqrt{\rho}(\lambda \I - \mu_{\x} |v_{\x}\rangle \langle v_{\x}| ) \sqrt{\rho} = P -N 
\eea
where $\mu_{\x} = \langle \Psi_{\x} |  \rho^{-1} | \Psi_{\x}\rangle$ and $|v_{\x}\rangle = \sqrt{\rho}^{-1} |\Psi_{\x}\rangle/\sqrt{\mu_{\x}}$. One can find that
\bea
P & = & \lambda  \sqrt{\rho}  (\I - |v_{\x}\rangle \langle v_{\x}| ) \sqrt{\rho} \nonumber \\
~\mathrm{and}~~ N & = &  ( \mu_{\x} -\lambda) \sqrt{\rho} |v_{\x} \rangle \langle v_{\x}|  \sqrt{\rho}  \nonumber 
\eea
where $\mu_{\x}\geq \lambda$. There are two solutions to MCMs. One is when $\lambda=\mu_\x$; consequently, $N=0$. In this case, we have $M_\x= a_\x \ketbra{\varphi_\x}$ where $\ket{\varphi_\x}$ is in Eq. \eqref{eq:glomcmanti} and $a_{\x}>0$. This measurement can realize unambiguous state identification, i.e., $\hat{C}_{\x,\max}^{(G)}=1$, for a state $\rho_{\x}$. Note that an outcome rate is bounded above $\eta_{\x} = \tr [\rho M_{\x} ]\leq 1/5$. 
 
The other is when $\lambda < \mu_{\x}$, for which we solve the optimality conditions in Eqs. (\ref{eq:ls}) and (\ref{eq:cs}) and have $M_\x= N_\x/ \tr[N_\x] = \ketbra{\nu(\lambda)}$ where
\bea
\ket{\nu(\lambda)} &\propto &(\sqrt{ \lambda^2+9}+3)\ket{\phi_{\x}}\otimes \ket{\phi_\x^\perp}+\lambda\ket{\phi_\x^\perp} \otimes \ket{\phi_\x} \nonumber\\
~\mathrm{where} ~~\lambda &=& \frac{2 \sqrt{3}(1-3 \eta_\x)}{\sqrt{(1-2 \eta_\x)(6 \eta_\x - 1)}}. \nonumber
\eea
In this case, we have $\eta_{\x} = \tr[\rho M_\x] > 1/5$. Recall that for $\eta_\x\geq 1/3$, an optimal POVM element is SEP. 

Let us define a gap $\Delta_{\x} = \hat{C}_{\x,\max}^{(G)} - \hat{C}_{\x,\max}^{(S)}$. Note that $\Delta_{\x}=0$ for $\eta_{\x} \geq 1/3$. With measurements in SEP, as shown in Eq. (\ref{eq:sepmcm}), we have $\hat{C}_{\x,\max}^{(S)} = 3/4$. Let us summarize the certification of NLWE according to an outcome rate: 
\begin{itemize}
    \item Range I: $\eta_{\x}\in (0,1/5]$,  $\Delta_{\x} = 1/4$, 
    \item Range II: $\eta_{\x}\in (1/5,1/3)$
    \bea
\Delta_{\x} = \frac{1}{8 \eta_\x} (1+\sqrt{3} \sqrt{8 \eta_\x - 12 \eta_\x^2 -1})-\frac{3}{4} \nonumber
\eea
    \item Range III: $\eta_{\x} \in [1/3,1]$,  $\Delta_{\x} = 0$.
\end{itemize}
The gap $\Delta_{\x}$ is plotted in Fig. \ref{fig:conc2}. Given outcome rates, NLWE of unknown measurements can be certified. 

{\it Certifying NLWE with inconclusive outcomes.} Let us consider cases where a certifiable confidence cannot be used to certify NLWE, i.e., $\hat{C}_\x \leq \hat{C}_{\x,\max}^{(S)}$ for all $\x$. We here show that inconclusive outcomes, alternatively to conclusive ones, can be used for the purpose. Let $\eta_0$ denote the rate of inconclusive outcome, which one may want to minimize. An outcome rate $\eta_0$ lower than what can be achieved by measurements in SEP can certify a measurement is GLOBAL. 

The problem for the minimum inconclusive outcome rate with SEP can be formulated as an SDP,
\bea \label{eq:incsdp}
\eta_{0,\min}^{(S)} &=& \mathrm{min}~ \tr[\rho M_0]\\
&& \mathrm{subject~to} ~
~~\hat{C}_{\x} \leq  \frac{q_{\x} \tr[\rho_{\x}M_{\x}]}{\tr[\rho M_{\x}]}  ~~\forall \x\nonumber \\
&&  M_0 + \sum_{\x=1}^n M_\x =\id,\nonumber\\
&& \{ M_0 \geq 0,~ M_\x \geq 0, ~\forall \x \} \subset \mathrm{SEP} \nonumber
\eea
where $\hat{C}_{\x}$ is a certifiable confidence from conclusive outcomes, which cannot be used to certify NLWE. The rate of inconclusive outcomes, lower than what can be achieved by measurements in SEP, i.e., $\eta_0 < \eta_{0,\min}^{(S)}$, certifies NLWE. 

We revisit antiparallel states in Eq. \eqref{eq:antitetraens} with a noisy POVM element in GLOBAL
\bea 
\widetilde{M}^{(G)}_\x=a ( \ketbra{\varphi_\x} + \gamma \id), \label{eq:mg}
\eea
where $a=(6/5 + 4 \gamma)^{-1}$ and $\ket{\varphi_\x}$ in Eq. (\ref{eq:glomcmanti}). Note also that for $\gamma\geq 1/10$, a confidence cannot be used for the certification, i.e.,
\bea
\hat{C}_{\x} = \frac{1}{4} \frac{\tr[\rho_\x \widetilde{M}^{(G)}_\x] }{\tr[\rho \widetilde{M}^{(G)}_\x]}\leq 
\frac{3}{4}.\nonumber
\eea
In this case, the inconclusive rate is given by,
\bea
\eta_{0} =1-\sum_{\x=1}^4 \tr[\rho \widetilde{M}_\x^{(G)}]=\frac{1}{3+10 \gamma},  \label{eq:eta0} 
\eea
which is less than or equal to $1/4$.  

From the optimization in Eq. (\ref{eq:incsdp}), we compute the minimum rate of inconclusive outcomes with SEP. It turns out that POVM elements in Eq. \eqref{eq:sepmcm} with $b_{\x}=1/2$ form an optimal measurement so that $\eta_{0,\min}^{(S)}=1/3$. The solution is obtained by analyzing the optimality conditions, see also the Supplemental Material. Therefore, the rate of inconclusive outcomes $\eta_0$ in Eq. (\ref{eq:eta0}) can certify NLWE, i.e., $\eta_{0,\min}^{(S)}=1/3>\eta_0$. 
 
Note also that there are cases where both conclusive and inconclusive outcomes can be used to certify NLWE. For the ensemble in Eq. \eqref{eq:antitetraens}, a square-root measurement, also known as a pretty-good measurement (PGM), given by $M_\x=\frac{1}{4}\sqrt{\rho}^{-1} \rho_\x \sqrt{\rho}^{-1}$, is not optimal for MCM. With this measurement, the confidence is given as $C_{\x} = (2+\sqrt{3})/4>3/4$ and the rate of inconclusive outcomes is $\eta_0 =0$. Both can be used to certify a gap between SEP and GLOBAL.

{\it Noise robustness in the certification of NLWE. } To demonstrate the noise robustness of the framework, we consider an ensemble of noisy antiparallel states, denoted by $\{\rho_{\x}\}_{\x=1}^4$ where each state is given as, 
\bea
\left(p|\phi_{\x}\rangle \langle \phi_{\x}| + \frac{(1-p)}{2}\I \right) \otimes \left(p|\phi_{\x}^{\perp}\rangle \langle \phi_{\x}^{\perp}| + \frac{(1-p)}{2}\I \right). \nonumber
\eea
From the optimality conditions in Eq. \eqref{eq:optimality}, an MCM is achieved by GLOBAL, $M_{\x}^{(G)}=a_\x \ketbra{\varphi_\x}$, where 
\bea
\ket{\varphi_\x} &=& \cos\theta \ket{\phi_\x}\otimes \ket{\phi_\x^\perp}+\sin\theta\ket{\phi_\x^\perp}\otimes \ket{\phi_\x}~\mathrm{and} \nonumber\\
\theta &=&\tan^{-1}\left(\frac{p(1+p)^2}{3p+p^2+\sqrt{9+7p^2-p^4+p^6}} \right) \nonumber
\eea
and note that $0<\theta\leq \tan^{-1}(\frac{1}{2})$. With the measurement, the MC for noisy antiparallel states is given by
\bea
C^{(G)}_{\x,\max}=\frac{1}{4(3-p^2)}\left(p^2+3+2p\sqrt{\frac{p^4-2p^2+9}{1+p^2}}\right). \label{eq:cg}
\eea
Note that $C^{(G)}_{\x,\max}$ becomes smaller as $p$ decreases. We compute the MC with SEP, see Eq. \eqref{eq:sepmcm}, as follows,
\bea
C_{\x, \max}^{(S)}=\frac{3}{4}\frac{(1+p)^2}{3+p^2} 
\eea
which is smaller than $C_{\x, \max}^{(G)}$ in Eq. (\ref{eq:cg}) for all $p$. Hence, NLWE can be demonstrated in the range of noise $p$. 

For the sDI certification framework, it turns out that $\hat{C}_{\x,\max}^{(G)} > \hat{C}^{(S)}_{\x, \max}$ only $\eta_\x < \eta_c$, where 
\bea
\eta_c=\frac{1}{12} \left(p^2 + 3 +\frac{3p(1-p^2)^2}{\sqrt{p^6-p^4-5 p^2+9}} \right). \nonumber
\eea 
That is, at a low detection rate, a larger gap can be found. For $\eta_\x \geq \bra{\varphi_1}\rho \ket{\varphi_1}$, the gap monotonically decreases until it happens $\eta_\x=\eta_c$, at which the gap vanishes, see Fig. \ref{fig:conc2}. Detailed derivations are shown in the Supplemental Material.

The certification NLWE with inconclusive outcomes is also robust against noise. Let us revisit a noisy POVM element, 
\bea \label{eq:noisyglobalpovm}
\widetilde{M}_\x^{(G)} = a (\ketbra{\varphi_\x}+\gamma \id)
\eea
where 
\bea
a=\begin{cases}
3( 12 \gamma+2+2\sin 2\theta)^{-1}, \text{ if } \theta \geq \frac{\pi}{12}\\
(4 \gamma + 2 - 2 \sin 2 \theta)^{-1}, \text{ if } \theta < \frac{\pi}{12}
\end{cases}
\eea
With this measurement, we have $\hat{C}_\x \leq C^{(S)}_{\x, \max}$ whenever
\bea
\gamma \geq \frac{1}{4}(1+p^2+ 2p \cos 2 \theta ) (1-\frac{C_{\x,\max}^{(S)}}{C_{\x,\max}^{(G)}})(4C_{\x,\max}^{(S)}-1)^{-1}.  \nonumber
\eea
The inconclusive rate by excluding all outcomes from the POVM elements $\{\widetilde{M}_\x^{(G)}\}$ is given by
\bea
\eta_{0}^{(G)} 
& =&1-a(1+4 \gamma + \frac{p^2}{3}(1-2 \sin 2 \theta)). \label{eq:etaGex}
\eea
The minimum inconclusive rate by measurements in SEP, i.e., $M_\x^{(S)}=\frac{1}{2}\ketbra{\phi_\x}\otimes \ketbra{\phi^\perp_\x}$, is given as
\bea
\eta_{0,\min}^{(S)}=\frac{1}{2}-\frac{p^2}{6}. 
\eea
In Fig. \ref{fig:placeholder}, we show that $\eta_{0,\min}^{(G)} < \eta_{0,\min}^{(S)}$. Thus, when conclusive outcomes do not demonstrate NLWE, i.e., $\hat{C}_{\x,\max }^{(S)} = \hat{C}_{\x}$, inconclusive outcomes can certify NLWE. 


\begin{figure}
    \centering
    \includegraphics[width=0.9\linewidth]{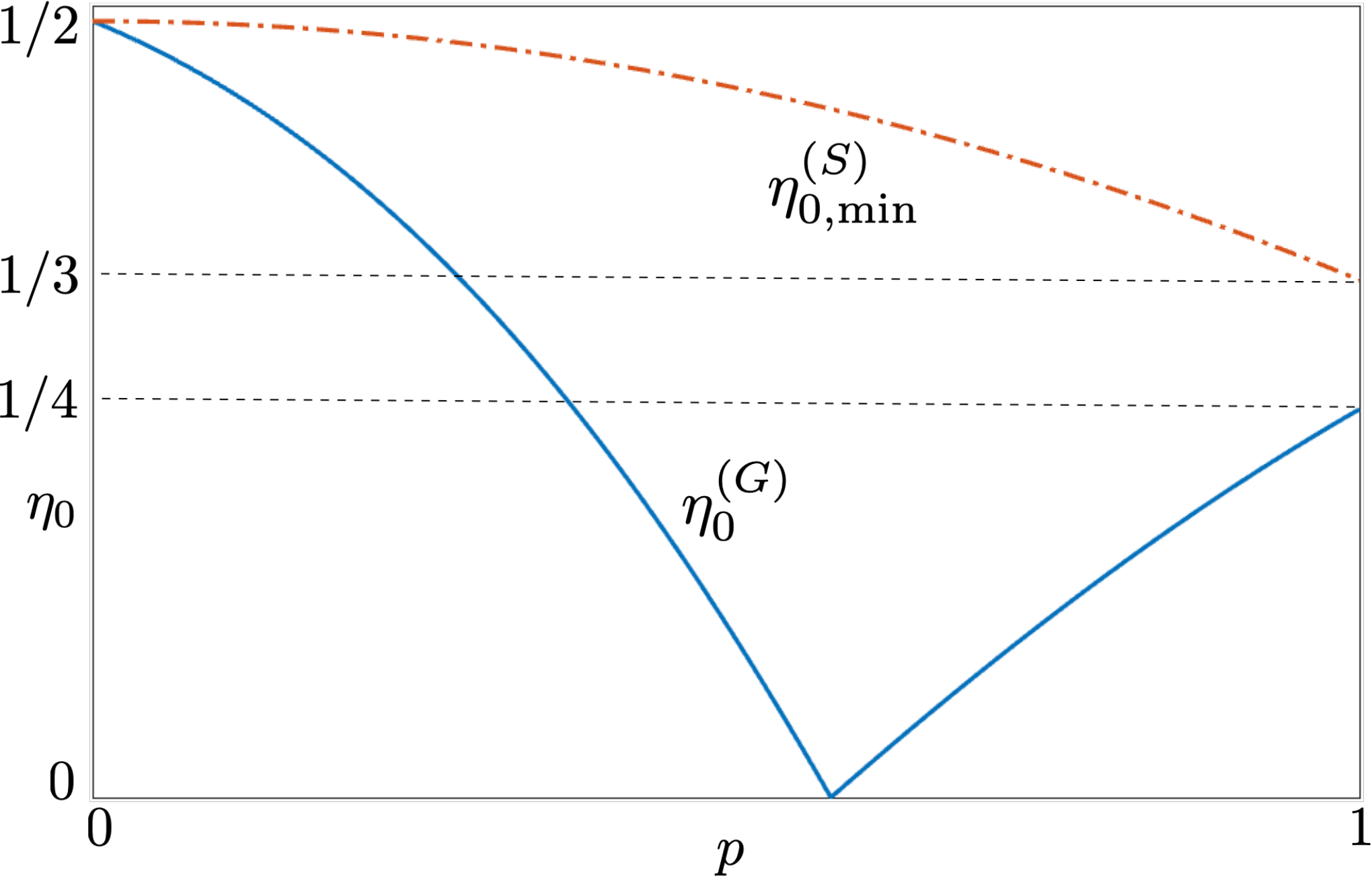}
    \caption{ The rate of inconclusive outcomes $\eta_{0}^{(G)}$ for a noisy POVM in Eq. \eqref{eq:noisyglobalpovm} (solid) is strictly lower than the rate by SEP (dotted) whenever $p>0$. Note that when $\theta=\pi/12$, at which $p \approx 0.58$, POVM elements $\{\widetilde{M}_\x^{(G)}\}_{\x=1}^4$ form a complete measurement, i.e., $\eta_0^{(G)}=0$.  }
    \label{fig:placeholder}
\end{figure}



{\it Conclusion.} In summary, we have presented the framework of demonstrating and certifying NLWE in terms of MCMs, where the confidence, i.e., a conditional probability {\it per} outcome, corresponds to a fine-grained guessing probability that unifies minimum-error and unambiguous discrimination strategies. Therefore, the framework can be used to detect a gap between GLOBAL and SEP in various state-discrimination-based quantum information applications. We have illustrated the framework with an ensemble of antiparallel quantum states, for which NLWE can be demonstrated and also certified. We also analyze the conditions on ensembles of states such that NLWE may exist. For instance, parallel states cannot show a gap between GLOBAL and SEP. 

We have shown that NLWE of unknown measurements can be certified from measurement outcomes, i.e., measurements that have not yet been verified. Then, given outcome rates, we determine whether a measurement for MC belongs to GLOBAL. Remarkably, inconclusive outcomes can also be used for certification purposes. All results, along with the demonstration and certification of NLWE, are robust to noisy state preparations. Our results, therefore, pave the way for verifying NWLE in realistic settings with present-day quantum technologies and enable its further applications. 

In future investigations, it would be interesting to devise NLWE-based practical applications, see also \cite{PhysRevX.15.021013, e21030325, hm9n-mkb3}, for instance, joint measurement and security of quantum communication against adversarial quantum repeaters. It is also interesting to unify various demonstrations of NLWE within the framework of MCMs where both conclusive and inconclusive outcomes can be exploited.

\section{acknowledgement}
H.L. acknowledges financial support from the Business Finland project BEQAH. J.B. is supported by the Institute for Information \& Communication Technology Promotion (IITP) (RS-2023-00229524, RS-2025-02304540, RS-2025-25464876, RS-2025-25464616).

\bibliographystyle{apsrev4-2}
\bibliography{reference}

\onecolumngrid

\appendix
\section*{ Supplemental Material}

\subsection*{I. The optimality conditions for maximum-confidence measurements}
In this section, we briefly review the optimality conditions for maximum-confidence measurements (MCMs).
For an ensemble of states $\{q_\x, \rho_\x\}_{\x=1}^n$, where $q_\x$ denotes \textit{a priori} probability, \textit{confidence} is defined as a conditional probability that a given measurement outcome $\x$ correctly identifies the state $\rho_\x$, i.e.
$C_\x=p_{P|M}(\x | \x)$, where $P$ and $M$ denote preparation and measurement respectively. 
The maximum confidence is derived from the following optimization,
\bea
C_{\x,\max}=\max_{M_\x \geq 0} \frac{q_\x \tr[\rho_\x M_\x]}{\tr[\rho M_\x]}, \nonumber
\eea
where $\rho=\sum_{\x} q_\x \rho_\x$ and $M_\x$ is a POVM element.
This problem can be formulated as a semi-definite program (SDP) by introducing the parameter $Q_\x=\frac{\sqrt{\rho} M_\x \sqrt{\rho}}{\tr[\rho M_\x]}$. The optimization becomes 
\bea
C_{\x,\max}=\max_{Q_\x \geq 0} \tr[\sqrt{\rho}^{-1} q_\x \rho_\x \sqrt{\rho}^{-1} Q_\x]:  \tr[Q_\x ]=1. \nonumber
\eea
It can be readily seen that the maximum confidence corresponds to the largest eigenvalue of $\sqrt{\rho}^{-1} q_\x \rho_x \sqrt{\rho}^{-1}$,
\bean
C_{\x, \max}=||\sqrt{\rho}^{-1} q_\x \rho_x \sqrt{\rho}^{-1}||_\infty
\eean
where $||\cdot||_\infty$ denotes an operator norm. 
The optimality conditions for an MCM \cite{lee2022maximum} read
\bean
C_{\x,\max} \rho - q_\x \rho_\x&=&r_\x \sigma_\x, \\
\tr[\sigma_\x M_\x]&=&0
\eean
where $r_\x >0$ and $\{\sigma_\x\}$ are quantum states called \textit{complementary} states. The first condition states that $\rho$ must be decomposed into $\rho_\x$ and $\sigma_\x$. Then, $M_\x$ is orthogonal to $\sigma_\x$.
Thus, solving the optimization reduces to finding the complementary states $\{\sigma_\x\}$. 

Let us derive the explicit condition for an MCM. We write the first condition above as
\bean
C_{\x,\max} \rho - q_\x \rho_\x &=& \sqrt{\rho}(C_{\x,\max} \id - \sqrt{\rho}^{-1} q_\x \rho_\x \sqrt{\rho}^{-1})\sqrt{\rho} \\
&=&r_\x \sigma_\x\nonumber.
\eean
Denote $d_\x$ the degeneracy of the largest eigenvalue of the operator $\sqrt{\rho}^{-1}q_\x \rho_\x \sqrt{\rho}^{-1}$ and $\ket{\lambda^{(j)}_{\x}}$ the associated eigenvectors, where $j=1,...,d_\x$.
Then, we see that the kernel of $\sigma_\x$ is spanned by the vectors $\sqrt{\rho}^{-1}\ket{\lambda_{\x}^{(j)}}$.  
 Define the subspace
\bea \label{eq:appmcmsubspace}
\mathcal{S}_{\x}^{MCM}=\mathrm{span} \{\sqrt{\rho}^{-1}\ket{\lambda_{\x}^{(1)}}, \ldots \sqrt{\rho}^{-1}\ket{\lambda_{\x}^{(d_\x)}} \}
\eea
which we call the \textit{MCM subspace}. Then, the necessary and sufficient condition of MCM is
\bea
\mathrm{supp} (M_\x) \subseteq \mathcal{S}^{MCM}_{\x} \nonumber.
\eea
We remark that $M_\x$ can always be written as a rank-one operator and, in general, has rank at most $d_\x$.

\subsection*{II. MCMs for noisy parallel and antiparallel states}
We consider ensembles of parallel and antiparallel symmetric, informationally complete (SIC) states, denoted by $S_{\|}=\{\ket{\phi_\x} \otimes \ket{\phi_\x} \}_{\x=1}^4$ and $S_{\perp}=\{\ket{\phi_\x}\otimes \ket{\phi_\x^\perp}\}_{\x=1}^4$, where 
\bea \label{eq:apptetra}
\ket{\phi_1}=\ket{0}, \ket{\phi_2}=\frac{1}{\sqrt{3}}\ket{0}+\sqrt{\frac{2}{3}}\ket{1},
\ket{\phi_{3,4}}=\frac{1}{\sqrt{3}}\ket{0}+e^{\pm \frac{2 \pi i}{3}}\sqrt{\frac{2}{3}}\ket{1}. \nonumber
\eea
with uniform \textit{a priori} probability $q_\x=\frac{1}{4}$ for all $\x$. We derive the global and separable MCMs for the parallel and antiparallel states subject to local depolarizing noise. We show that MCMs do not find any gap between GLOBAL and SEP for the noisy parallel states. In contrast, for noisy antiparallel states, MCMs feature NLWE for any amount of white noise.  

\subsubsection{Noisy parallel states}
Assume that the ensemble of parallel states $S_{\|}=\{\ket{\phi_\x}^{\otimes 2} \}_{\x=1}^4$ undergoes a local depolarizing noise. The resulting states are described as
\bea
\rho_{\x}=\left(p|\phi_{\x}\rangle \langle \phi_{\x}| + \frac{(1-p)}{2}\I \right) \otimes \left(p|\phi_{\x}^{}\rangle \langle \phi_{\x}^{}| + \frac{(1-p)}{2}\I \right) \nonumber
\eea
where $p \in (0,1]$. The ensemble state is
\bean
\rho=\frac{1}{4}\sum_{\x=1}^4 \rho_\x=\frac{3+p^2}{12}\Pi_{\mathrm{sym}}+ \frac{1-p^2}{4}\ketbra{\psi^-}
\eean
where $\ket{\psi^{-}}=\frac{1}{\sqrt{2}}(\ket{01}-\ket{10})$ and $\Pi_{\mathrm{sym}}$ is the projector onto the symmetric subspace. To derive the global MCM, we need to find the largest eigenvalue and its associated eigenvector of the operator $\sqrt{\rho}^{-1} \rho_\x\sqrt{\rho}^{-1}$, which admits the spectral decomposition 
\bean
\sqrt{\rho}^{-1} \rho_\x\sqrt{\rho}^{-1}=
\frac{3}{3+p^2} \Big[(1+p)^2\ketbra{\phi_\x}^{\otimes 2}+(1-p)^2\ketbra{\phi^\perp_\x}^{\otimes 2}+(1-p^2)\ketbra{\phi_\x^+} \Big] +\ketbra{\psi^-}
\eean
where $\ket{\phi_\x^+}=\frac{1}{\sqrt{2}}(\ket{\phi_\x}\otimes \ket{\phi_\x^\perp}+\ket{\phi_\x^\perp}\otimes \ket{\phi_\x})$. Its largest eigenvalue is $\frac{3(1+p)^2}{3+p^2}$ with the corresponding eigenvector given as $\ket{\phi_\x}\otimes\ket{\phi_\x}$. 
Therefore, the MCM subspace $\mathcal{S}^{MCM}_{\x}$ in Eq. \eqref{eq:appmcmsubspace} is spanned by a vector $\sqrt{\rho}^{-1} \ket{\phi_\x}\ket{\phi_\x} \propto \ket{\phi_\x}\ket{\phi_\x}$, and 
the global MCM is a separable measurement,
\bean
M_\x^{(G)}= a_\x \ketbra{\phi_\x}\otimes \ketbra{\phi_\x}
\eean
where $a_\x \in (0, 1]$. This measurement yields the confidence
\bea \label{eq:appglomcparallel}
C_{\x,\max}^{(G)}=\frac{3(1+p)^2}{4(3+p^2)}.
\eea
We conclude that for the parallel states with local white noise, MCMs do not find the gap between GLOBAL and SEP.

\subsubsection{ Noisy antiparallel  states}
We consider noisy antiparallel states,
\bea \label{eq:appnoisyanti}
\rho_{\x}=\left(p|\phi_{\x}\rangle \langle \phi_{\x}| + \frac{(1-p)}{2}\I \right) \otimes \left(p|\phi_{\x}^{\perp}\rangle \langle \phi_{\x}^{\perp}| + \frac{(1-p)}{2}\I \right)
\eea
where $p \in (0,1]$. The ensemble state is
\bean
\rho=\frac{1}{4}\sum_{\x=1}^4 \rho_\x=\frac{3-p^2}{12}\Pi_{\mathrm{sym}}+\frac{1+p^2}{4}\ketbra{\psi^-}
\eean
With a bit of algebra, we find that the MCM subspace $\mathcal{S}^{MCM}_{\x}$ is spanned by an entangled state 
\bea \label{app:globalmcm}
\ket{\varphi_\x}= \cos\theta \ket{\phi_\x}\otimes \ket{\phi_\x^\perp}+\sin\theta\ket{\phi_\x^\perp}\otimes \ket{\phi_\x}.
\eea
where
\bea \label{eq:apptheta}
\theta=\tan^{-1}\left(\frac{p(1+p)^2}{3p+p^2+\sqrt{9+7p^2-p^4+p^6}}\right). 
\eea
We remark that $0<\theta\leq \tan^{-1}(\frac{1}{2})$ and the upper bound is attained when $p=1$. Therefore, the global MCM is an entangled measurement, 
\bea
M_\x^{(G)}=a_\x \ketbra{\varphi_\x}, \nonumber
\eea
where $a_\x \in (0,1]$ can be freely chosen. This measurement yields the maximum confidence
\bea
C^{(G)}_{\x,\max}=\frac{1}{4(3-p^2)}\left(p^2+3+2p\sqrt{\frac{p^4-2p^2+9}{1+p^2}}\right). \nonumber
\eea
Since $M^{(G)}_\x$ is entangled, the MCM for noisy antiparallel states features NLWE for all $p \in (0,1]$.
\begin{figure}
    \centering
    \includegraphics[width=0.5\linewidth]{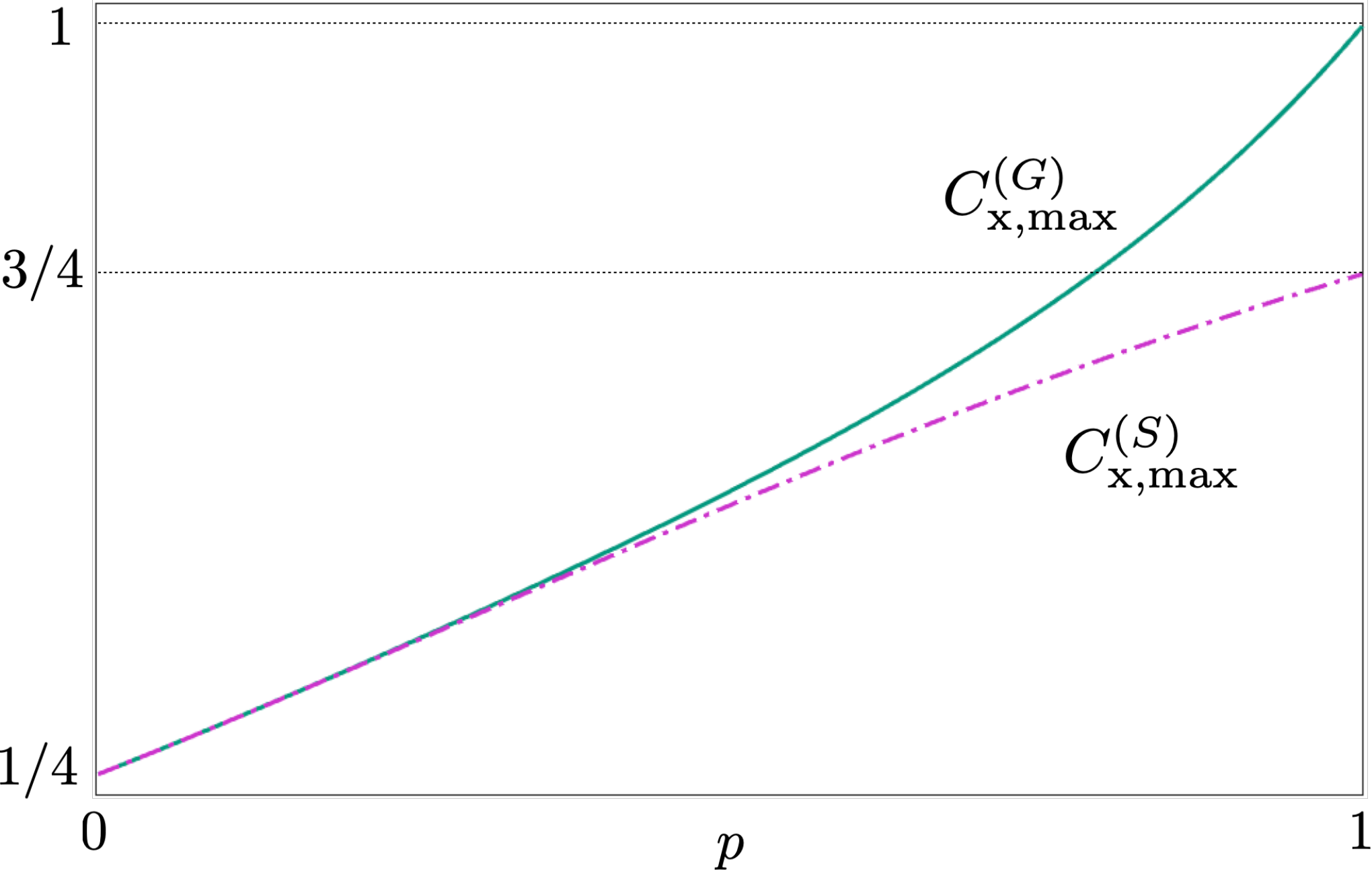}
    \caption{The maximum confidence for noisy antiparallel states in Eq. \eqref{eq:appnoisyanti} by GLOBAL (solid) and SEP (dotted) is shown. }
    \label{fig:appfig}
\end{figure}
When $p=1$, the entangled state reduces to
\bea
\ket{\varphi_\x}=\frac{1}{\sqrt{5}} (2\ket{\phi_\x}\otimes\ket{\phi_\x^\perp}+\ket{\phi_\x^\perp}\otimes\ket{\phi_\x}). \label{eq:appud}
\eea
Since $\langle \Psi_\y| \varphi_\x\rangle =0$ for $\y \neq \x$, a POVM element $M_\x^{(G)}=a_\x \ketbra{\varphi_\x}$ excludes all states $\ket{\Psi_\y}$ except $\y=\x$. 
Therefore, the global MCM unambiguously discriminates the states, achieving confidence
\bea
C_{\x,\max}^{(G)}=1, \forall \x. \nonumber
\eea

Let us turn to finding the separable MCM for the noisy antiparallel states. The optimization is given as
\bea
C_{\x,\max}^{(S)}=\max_{M_\x \in \mathrm{SEP}} \frac{1}{4} \frac{ \tr[\rho _\x M_\x]}{\tr[\rho M_\x]} \nonumber
\eea
where $M_\x$ is now restricted to be a separable operator. We consider a rank-one separable operator $M_\x=\ketbra{e_\x}\otimes \ketbra{f_\x}$ as an ansatz. 
Using this ansatz, we solve the optimization,
\bea \label{eq:appsepmc}
C_{\x,\max}^{(S)}&=&
\max_{\ket{e_\x}, \ket{f_\x} } \frac{1}{4}\frac{\tr[\rho_\x \ketbra{e_\x}\otimes \ketbra{f_\x}]}{\tr[\rho \ketbra{e_\x}\otimes \ketbra{f_\x}]} \nonumber\\
&=&\max_{\ket{e_\x},\ket{f_\x}} \frac{(p|\ip{\phi_\x}{e_\x}|^2 + \frac{1-p}{2})(p|\ip{\phi_\x^\perp}{f_\x}|^2 + \frac{1-p}{2})}{\sum_\y (p|\ip{\phi_\y}{e_\x}|^2 + \frac{1-p}{2})(p|\ip{\phi_\y^\perp}{f_\x}|^2 + \frac{1-p}{2})} \nonumber\\
&=&\max_{\ket{e_\x},\ket{f_\x^\perp}} \frac{(p|\ip{\phi_\x}{e_\x}|^2 + \frac{1-p}{2})(p|\ip{\phi_\x}{f_\x^\perp}|^2 + \frac{1-p}{2})}{\sum_\y (p|\ip{\phi_\y}{e_\x}|^2 + \frac{1-p}{2})(p|\ip{\phi_\y}{f_\x^\perp}|^2 + \frac{1-p}{2})} \nonumber\\
&=&\frac{3}{4}\frac{(1+p)^2}{3+p^2}
\eea
where we used the relation $|\ip{\phi^\perp_\y}{f_\x}|=|\ip{\phi_\y}{f_\x^\perp}|$ and that the optimization is equivalent to finding the MCM for the parallel states in Eq. \eqref{eq:appglomcparallel}. Then, the separable MCM is given by
\bea \label{eq:appsepmcm}
M_\x^{(S)}=b_\x \ketbra{\phi_\x}\otimes\ketbra{\phi_\x^\perp}
\eea
where $b_\x \in (0, 1]$ can be freely chosen. Theorefore, when the measurement is restricted to be separable, the optimal measurement is in the direction of the states.

\subsection*{III. The optimality conditions for the certifiable maximum confidence }

In this section, we derive the optimality conditions for the certifiable maximum confidence  under given outcome rates. Then, we provide solutions to the optimality conditions. 

Consider an ensemble of quantum states $S=\{q_\x \rho_\x \}_{\x=1}^n$. Let $\{\eta_\x\}_{\x=1}^n$ denote outcome rates of the experiment, i.e. $\eta_\x=\tr[\rho M_\x]$ where $\rho=\sum_\x q_\x\rho_\x$ and $M_\x$ is a POVM element for outcome $\x$ that describe the measurement in the experiment. Our aim is to find the maximally achievable confidence by GLOBAL and SEP that is compatible with the experimental statistics, especially outcome rates.  Then, with the trusted quantum states, the untrusted measurement is certfied to feature NLWE if the confidence it produces, called certifiable confidence $\{\hat{C}_\x\}_{\x=1}^n$, cannot be achieved by SEP. 

The problem of certifiable maximum confidence can be formulated as an SDP,
\bea
&&\hat{C}_{\x, \max}=\max \frac{q_\x}{\eta_\x}\tr[\rho_\x M_\x] \nonumber\\
&&\textrm{subject~to~} \tr[\rho M_\x]=\eta_\x, 0\leq M_\x\leq \id, \nonumber
\eea
We denote $\hat{C}^{(G)}_{\x, \max}$ and $\hat{C}^{(S)}_{\x, \max}$ the certifiable maximum confidence by GLOBAL and SEP, respectively. Let us derive the dual problem. For convenience, we first omit the prefactor $\frac{q_\x}{\eta_\x}$.  Introducing the dual parameters $P,N \geq 0$ for the inequality constraints and $\lambda \in \mathbb{R}$ for the equality constraint, the Lagrangian reads
\bean
\mathcal{L}(M_\x, P, N,\lambda)&=&\tr[\rho_\x M_\x] +\tr[M_\x P] + \tr[(\id- M_\x) N]+ \lambda(\eta_\x-\tr[\rho M_\x]) \nonumber\\
&=&\tr[M_\x (\rho_\x+ P - N -\lambda \rho)] +\tr[N]+\lambda \eta_\x. \nonumber
\eean
The dual function is derived as
\bean
g(P,N,\lambda)&=&\sup_{M_\x} \mathcal{L}(M_\x, P, N, \lambda)\\
&=&\begin{cases}
    \tr[N]+\lambda \eta_\x, \textrm{ if }  \rho_\x+P-N-\lambda \rho =0\\
    \infty, \textrm{ otherwise }
\end{cases}
\eean
The dual function does not diverge only if $\rho_\x+P-N-\lambda \rho=0$, which leads to the dual problem 
\bea
\hat{C}_{\x,\max} &=& \mathrm{min~} \tr[N] + \lambda \eta_{\x} \nonumber\\
&& \mathrm{subject~to}~~ N \geq 0, N+\lambda \rho \geq \rho_{\x}. \nonumber
\eea
Strong duality holds since a strictly feasible point exists (e.g., $M_\x=\eta_\x \id$), and therefore the primal and the dual problems yield the same optimal value. 
Bringing back the prefactor $\frac{q_\x}{\eta_\x}$, the certifiable maximum confidence is derived as
\bea \label{eq:appcertmc}
\hat{C}_{\x,\max}^{(G)}=q_\x \left(\lambda+\frac{\tr[N]}{\eta_\x} \right)\, .
\eea
Besides the constraints in the primal and the dual problems, the optimality conditions are listed as
\bea \label{eq:appkkt}
\textrm{(Lagrangian stability) } P-N &=&\lambda \rho - \rho_\x \\
\textrm{(Complementary slackness) } M_\x P &=&0, (\id - M_\x) N =0. \nonumber
\eea

Three observations are immediately made from these optimality conditions. 
First, these conditions imply that $P$ and $N$ correspond to the positive and negative parts of $\lambda \rho-\rho_\x$, respectively. Second, $\lambda$ can be bounded between the maximum and minimum eigenvalues of the operator $\sqrt{\rho}^{-1} \rho_\x \sqrt{\rho}^{-1}$; otherwise, $M_\x$ takes only trivial solutions, $M_\x=0$ or $M_\x=\id$. 
Lastly, since $M_\x N=N$,  $M_\x$ must act as an identity to $N$. To be specific, let us denote the spectral decomposition of $N$ as $N=\sum n_i \ketbra{n_i}$. Then, $M_\x$ must take the form
\bea
M_\x=\sum_i \ketbra{n_i}+L \nonumber
\eea 
where $L \geq 0$ is orthogonal to $N$.

\subsection*{IV. The certifiable maximum confidence for the antiparallel states}
\label{app:d}

We consider an ensemble of noiseless antiparallel states, $S_{\perp}=\{ \ket{\Psi_\x}\}_{\x=1}^4$ where 
\bea
\ket{\Psi_\x}=\ket{\phi_\x} \otimes \ket{\phi_\x^\perp} \nonumber
\eea
and $\{\ket{\phi_\x}\}_{\x=1}^4$ are the SIC states given in Eq. \eqref{eq:apptetra}. We find the certifiable maximum confidence for $S_{\perp}$ by GLOBAL and SEP measurements. For that purpose, we solve the optimality conditions in Eq. \eqref{eq:appkkt} by finding suitable parameters $\lambda, P, N$ and $M_\x$ that satisfy the conditions. 
Let us write the optimality condition as
\bea \label{eq:appopt1}
\sqrt{\rho}(\lambda \I - \mu_{\x} |v_{\x}\rangle \langle v_{\x}| ) \sqrt{\rho} = P -N 
\eea
where $\mu_{\x} = \langle \Psi_{\x} |  \rho^{-1} | \Psi_{\x}\rangle=4$ and $|v_{\x}\rangle = \sqrt{\rho}^{-1} |\Psi_{\x}\rangle/\sqrt{\mu_{\x}}$. Note that $\lambda$ can be bounded in the interval $\lambda \in [0,\mu_\x]$. This is because if it takes other values, then $P$ or $N$ are full-rank in the support of $\rho$, and therefore only trivial solutions $M_\x=0$ or $M_\x=\id$ exist. 

Since $P$ and $N$ are positive and negative parts of $\lambda \rho - \rho_\x$, let us find the spectral decomposition of $\lambda\rho-\rho_\x$,
\bea \label{eq:appopt2}
P-N&=&\lambda \rho - \rho_\x \nonumber\\
&=&\sum_{i=1}^4 \nu_i(\lambda) \ketbra{\nu_i(\lambda)}
\eea
where the eigenvalues are
\bean
&&\nu_1(\lambda)=\frac{1}{6} (2\lambda -3-\sqrt{\lambda^2 +9})\\
&&\nu_2(\lambda)=\nu_3(\lambda)=\frac{\lambda}{6}\\
&&\nu_4(\lambda)=\frac{1}{6} (2\lambda -3+\sqrt{\lambda^2 +9})
\eean
and the corresponding eigenvectors are given as
\bean
&&\ket{\nu_1(\lambda)}\propto(\sqrt{\lambda^2+9}+3)\ket{\phi_\x} \otimes \ket{\phi_\x^\perp}+\lambda  \ket{\phi_\x^\perp} \otimes \ket{\phi_\x}\\
&&\ket{\nu_2(\lambda)}=\ket{\phi_\x} \otimes \ket{\phi_\x}\\
&&\ket{\nu_3(\lambda)}=\ket{\phi_\x^\perp} \otimes \ket{\phi_\x^\perp}\\
&&\ket{\nu_4(\lambda)}\propto(\sqrt{\lambda^2 p^2+9}-3)\ket{\phi_\x}\otimes \ket{\phi_\x^\perp}-\lambda p \ket{\phi_\x^\perp}\otimes \ket{\phi_\x}.
\eean
Recall that we only consider $\lambda \in [0,4]$; otherwise, $M_\x$ has only trivial solutions. Our approach is to find all suitable parameters $P,N$, and $M_\x$ for each possible case of $\lambda$: $\lambda=4$, $0<\lambda<4$, and $\lambda=0$ such that they satisfy the optimality conditions. This approach is called \textit{complementary problem} \cite{flatt2022contextual}.  
\subsubsection{Solution 1: $\lambda=4$}
When $\lambda=4$, we have $\nu_1(\lambda)=0$ and $\nu_i(\lambda)>0$ for $i=2,3,4$. Therefore, we can characterize the positive and negative part of $\lambda \rho -\rho_\x$ as
\bea
&&P=\sum_{i=2}^4 \nu_i (\lambda) \ketbra{\nu_i(\lambda)},\nonumber\\
&&N=0.\nonumber
\eea
To find $M_\x$ that satisfies the complementary slackness conditions in Eq. \eqref{eq:appkkt}, we write $P$ as
\bea
P=4\rho - \rho_\x=\sqrt{\rho}(4 \id - \sqrt{\rho}^{-1} \ketbra{\Psi_\x} \sqrt{\rho}^{-1})\sqrt{\rho} \nonumber
\eea
Since $\langle \Psi_\x | \rho^{-1} | \Psi_\x \rangle=4$, we have
\bea
M_\x=a_\x \ketbra{\varphi_\x}\nonumber
\eea
where $a_\x \in (0,1]$ and
\bea
\ket{\varphi_\x}=\frac{\rho^{-1} \ket{\Psi_\x}}{||\rho^{-1} \ket{\Psi_\x}||}=\frac{1}{\sqrt{5}}(2 \ket{\phi_\x} \otimes \ket{\phi_\x^\perp} + \ket{\phi_\x^\perp} \otimes \ket{\phi_\x}).\nonumber
\eea
Note that $M_\x$ unambiguously identifes the states, see Eq. \eqref{eq:appud}. 
Lastly, we can find the region of outcome rate such that this solution is valid by solving $\tr[\rho M_\x]=\eta_\x$. Since $a_\x \in (0,1]$, the range for the outcome rate is bounded by $\eta_\x\leq \langle \varphi_\x | \rho | \varphi_\x \rangle=\frac{1}{5}$. We have solved the certifiable maximum confidence by GLOBAL for the low outcome rate region,
\bea
\hat{C}_{\x,\max}^{(G)}=1, \text{for } \eta_\x \in (0,1/5]\nonumber
\eea
We will later show that the gap between GLOBAL and SEP is maximal in this range of outcome rate. 
\subsubsection{Solution 2: $\lambda \in (0,4)$}
Let us consider the second case, $0< \lambda < 4$. Note that the eigenvalues $\nu_i(\lambda)$ are monotonically increasing in $\lambda$ for all $i$. We find that, when $0<\lambda<4$, $\nu_1(\lambda)<0$ and $\nu_i(\lambda)>0$ for $i=2,3,4$. Accordingly, $P$ and $N$ are given as
\bea
&&P=\sum_{i=2}^4 \nu_i(\lambda) \ketbra{\nu_i(\lambda)}\nonumber\\
&&N=-\nu_1(\lambda) \ketbra{\nu_1(\lambda)}\nonumber
\eea
To satisfy the complementary slackness conditions, $M_\x$ must be
\bea
M_\x=\ketbra{\nu_1(\lambda)}\nonumber.
\eea
Note that $\ket{\nu_1(\lambda)}=\ket{\varphi_\x}$ when $\lambda=4$ and $\ket{\nu_1(\lambda)}=\ket{\Psi_\x}$ when $\lambda=0$. The state $\ket{\nu_1(\lambda)}$ is entangled for $0<\lambda<4$. Solving $\tr[\rho M_\x]=\eta_\x$, we find the relation
\bean
\lambda=\frac{2 \sqrt{3}(1-3 \eta_\x)}{\sqrt{(1-2 \eta_\x)(6 \eta_\x-1)}}
\eean
Therefore, the range $ \lambda \in (0,4)$ corresponds to the range of the outcome rate
\bea
\eta_\x \in (1/5, 1/3)\nonumber
\eea
for which $M_\x$ must be entangled. 
From Eq. \eqref{eq:appcertmc}, we find the certifiable maximum confidence by GLOBAL measurements,
\bea
\hat{C}_{\x, \max}^{(G)}&=&\frac{1}{4}(\lambda + \frac{\tr[N]}{\eta_\x})\nonumber\\
&=& \frac{1}{8 \eta_\x} (1+\sqrt{3} \sqrt{8 \eta_\x - 12 \eta_\x^2 -1})\nonumber
\eea
\subsubsection{Solution 3: $\lambda=0$}
Our last case is when $\lambda=0$, in which we have $\nu_1(\lambda)<0$ and $\nu_i(\lambda)=0$ for $i=2,3,4$. It follows that $P$ and $N$ are
\bea
P=0, N=\ketbra{\Psi_\x}.\nonumber
\eea
A separable POVM element,
\bea
M_\x=\alpha_\x \ketbra{\Psi_\x}+ (1-\alpha_\x) \id,\nonumber
\eea
where $\alpha_\x \in [0,1]$, satisfies the complementary slackness conditions. The range for $\eta_\x$ for which this solution is valid is then
\bea
\eta_\x \in [1/3, 1].\nonumber
\eea
The certifiable maximum confidence in Eq. \eqref{eq:appcertmc} is 
\bea
\hat{C}_{\x, \max}^{(G)}=\frac{1}{4\eta_\x}\nonumber.
\eea
Since $M_\x$ can be separable, NLWE cannot be certified in this range of outcome rate. 

\subsubsection{The certifiable maximum confidence by SEP measurements}
In the above, we showed that the outcome rate must be $\eta_\x<\frac{1}{3}$ to certify NLWE. Let us find the certifiable maximum confidence by SEP. Recall that the separable MCM in Eq. \eqref{eq:appsepmcm} is 
\bea
M_\x^{(S)}=b_\x \ketbra{\phi_\x}\otimes \ketbra{\phi_\x^\perp}\nonumber
\eea
where $b_\x \in (0,1]$. This measurement yields confidence $C_{\x, \max}^{(S)}=3/4$ upto the outcome rate $\eta_\x\leq\langle \Psi_\x | \rho | \Psi_\x\rangle =1/3$. Therefore, the certifiable maximum confidence by SEP measurements is
\bea
\hat{C}_{\x, \max}^{(S)}=3/4\nonumber
\eea
in the range of $\eta_\x \in (0,1/3]$. The gap $\Delta_\x=\hat{C}^{(G)}_{\x,\max}-\hat{C}^{(S)}_{\x,\max}$ in the three ranges of outcome rate is given by
\begin{itemize}
    \item  Range I: $\eta_{\x}\in (0,1/5]$,  $\Delta_{\x} = 1/4$, 
    \item Range II: $\eta_{\x}\in (1/5,1/3)$
    $
\Delta_{\x} = \frac{1}{8 \eta_\x} (1+\sqrt{3} \sqrt{8 \eta_\x - 12 \eta_\x^2 -1})-\frac{3}{4} \nonumber
$
    \item Range III: $\eta_{\x} \in [1/3,1/5]$,  $\Delta_{\x} = 0$
\end{itemize}
The gap is maximal in Range I, decreases monotonically throughout Range II, and vanishes in Range III. 

\subsection*{ V. Certifiable maximum confidence for noisy antiparallel states}
We consider the noisy antiparallel states that are subject to local depolarizing noise,
\bea
\rho_{\x}=\left(p|\phi_{\x}\rangle \langle \phi_{\x}| + \frac{(1-p)}{2}\I \right) \otimes \left(p|\phi_{\x}^{\perp}\rangle \langle \phi_{\x}^{\perp}| + \frac{(1-p)}{2}\I \right). \nonumber
\eea
We show that NLWE can be certified from conclusive outcome $\x$ if and only if $\eta_\x < \eta_c$, where
\bea
\eta_c=\frac{1}{12}(3+p^2+\frac{3p(1-p^2)^2}{\sqrt{p^6-p^4-5 p^2+9}}). \nonumber
\eea

Let us analyze the optimality conditions in Eq. \eqref{eq:appkkt}. To find $P$ and $N$, we first derive the spectral decomposition of $\lambda \rho - \rho_\x$,
\bea
P-N&=&\lambda \rho - \rho_\x\nonumber \\
&=&\sum_{i=1}^4 \nu_i(\lambda) \ketbra{\nu_i(\lambda)}\nonumber
\eea
where the eigenvalues are
\bean
&&\nu_1(\lambda)=\frac{1}{12} \big(\lambda (p^2+3)-3(1+p^2)-2p\sqrt{\lambda^2 p^2+9} \big)\\
&&\nu_2(\lambda)=\nu_3(\lambda)=\frac{1}{12} \big(\lambda(3-p^2)-3(1-p^2) \big)\\
&&\nu_4(\lambda)=\frac{1}{12} \big(\lambda (p^2+3)-3(1+p^2)+2p\sqrt{\lambda^2 p^2+9} \big)
\eean
and the corresponding eigenvectors are given as
\bean
&&\ket{\nu_1(\lambda)}\propto(\sqrt{\lambda^2 p^2+9}+3)\ket{\phi_\x}\otimes \ket{\phi_\x^\perp}+\lambda p \ket{\phi_\x^\perp}\otimes \ket{\phi_\x}\\
&&\ket{\nu_2(\lambda)}=\ket{\phi_\x}\otimes \ket{\phi_\x}\\
&&\ket{\nu_3(\lambda)}=\ket{\phi_\x^\perp}\otimes \ket{\phi_\x^\perp}\\
&&\ket{\nu_4(\lambda)}\propto(\sqrt{\lambda^2 p^2+9}-3)\ket{\phi_\x}\otimes \ket{\phi_\x^\perp}-\lambda p \ket{\phi_\x^\perp}\otimes \ket{\phi_\x}. 
\eean
To find $P$ and $N$ in terms of $\lambda$, we need to determine the signs of $\nu_i(\lambda)$. For that purpose, we rewrite $\lambda \rho - \rho_\x$ as
\bea
\lambda \rho - \rho_\x=\sqrt{\rho}(\lambda \id - \sqrt{\rho}^{-1} \rho_\x \sqrt{\rho}^{-1} )\sqrt{\rho}.\nonumber
\eea
Since $\lambda \rho - \rho_\x$ and $\lambda \id -  \sqrt{\rho}^{-1} \rho_\x \sqrt{\rho}^{-1}$ have the same rank, it follows that $\nu_i(\lambda)$ vanishes when $\lambda$ corresponds to one of the eigenvalues of $\sqrt{\rho}^{-1} \rho_\x \sqrt{\rho}^{-1}$. Let $\{\lambda_i\}_{i=1}^4$ denote the eigenvalues of $\sqrt{\rho}^{-1} \rho_\x \sqrt{\rho}^{-1}$,
\bean
&&\lambda_1=\frac{1}{3-p^2}\left(p^2+3+2p\sqrt{\frac{p^4-2p^2+9}{1+p^2}}\right)\\
&&\lambda_2=\lambda_3=\frac{3-3p^2}{3-p^2}\\
&&\lambda_4=\frac{1}{3-p^2}\left(p^2+3-2p\sqrt{\frac{p^4-2p^2+9}{1+p^2}}\right).
\eean
which are labeled in decreasing order, $\lambda_1\geq \lambda_2 \geq \lambda_4$. It can be checked that $\nu_i(\lambda_i)=0$ for all $i$. Moreover, $\nu_i(\lambda)$ is monotonally increasing in $\lambda$ for all $i$. Therefore, we make the following remark,
\bea
&&\nu_i(\lambda_j) \geq 0, \text{ if } \lambda_j\geq \lambda_i\nonumber\\
&&\nu_i(\lambda_j) < 0, \text{ if } \lambda_j< \lambda_i.\nonumber
\eea
These relationships determine $P$ and $N$ in terms of $\lambda$. Recall that $\lambda \in [\lambda_4, \lambda_1]$. Then, $P$ and $N$ take five different forms depending on $\lambda$, listed below.
\begin{itemize}
\item For $\lambda=\lambda_1$, \bea
P=\sum_{i=2}^4 \nu_i(\lambda)\ketbra{\nu_i(\lambda)}, ~N=0. \nonumber
\eea
\item For $\lambda_2 < \lambda < \lambda_1$,
\bean
&&P=\sum_{i=2}^4 \nu_i(\lambda)\ketbra{\nu_i(\lambda)},~ N=-\nu_1(\lambda)\ketbra{\nu_1(\lambda)}.
\eean
\item For $\lambda=\lambda_2$,
\bean
P=\nu_4(\lambda) \ketbra{\nu_4(\lambda)}, ~N=-\nu_1(\lambda)\ketbra{\nu_1(\lambda)}.
\eean
\item For $\lambda_4 < \lambda < \lambda_2$, 
\bean
&&P=\nu_4(\lambda) \ketbra{\nu_4(\lambda)}, ~N=-\sum_{i=1}^{3}\nu_i(\lambda)\ketbra{\nu_i(\lambda)}
\eean
\item For $\lambda= \lambda_4$, 
\bean
P=0, ~N=-\sum_{i=1}^{3}\nu_i(\lambda)\ketbra{\nu_i(\lambda)}
\eean
\end{itemize}
The complementary slackness conditions in Eq. \eqref{eq:appkkt}, i.e., $P M_\x=0$ and $N(\id - M_\x)=0$, lead to the following solutions of $M_\x$,
\bean
M_\x=
\begin{cases}
    a_\x \ketbra{\nu_1(\lambda)}, \text{ for }\lambda=\lambda_1\\
    \ketbra{\nu_1 (\lambda)}, \text{ for }\lambda_2 < \lambda <\lambda_1\\
    \ketbra{\nu_1(\lambda)} + L, \text{ for }\lambda=\lambda_2\\
    \id-\ketbra{\nu_3(\lambda)}, \text{ for }\lambda_4<  \lambda <\lambda_2\\
   \id-b_\x \ketbra{\nu_3(\lambda)}, \text{ for }\lambda=\lambda_4
\end{cases}
\eean
where $a_\x,b_\x\in [0,1]$ and $L\geq0$ is an operator such that  $\mathrm{supp}(L) \subseteq \mathrm{span} \{ \ket{\nu_2 (\lambda_2)},\ket{\nu_3 (\lambda_2)} \}$.

Since $\ket{\nu_1(\lambda)}$ is an entangled state for $\lambda_2<\lambda\leq \lambda_1$, NLWE can be certified for outcome rates satisfying $\eta_\x=\tr[\rho M_\x]$. On the other hand, for $\lambda_3 \leq \lambda < \lambda_2$, the POVM element $M_\x$ is shown to be separable by checking the positivity of its partial transpose. Therefore, the critical outcome rate $\eta_c$ at which NLWE can no longer be certified must occur when $\lambda=\lambda_2$. 

\begin{figure}
    \centering
    \includegraphics[width=0.5\linewidth]{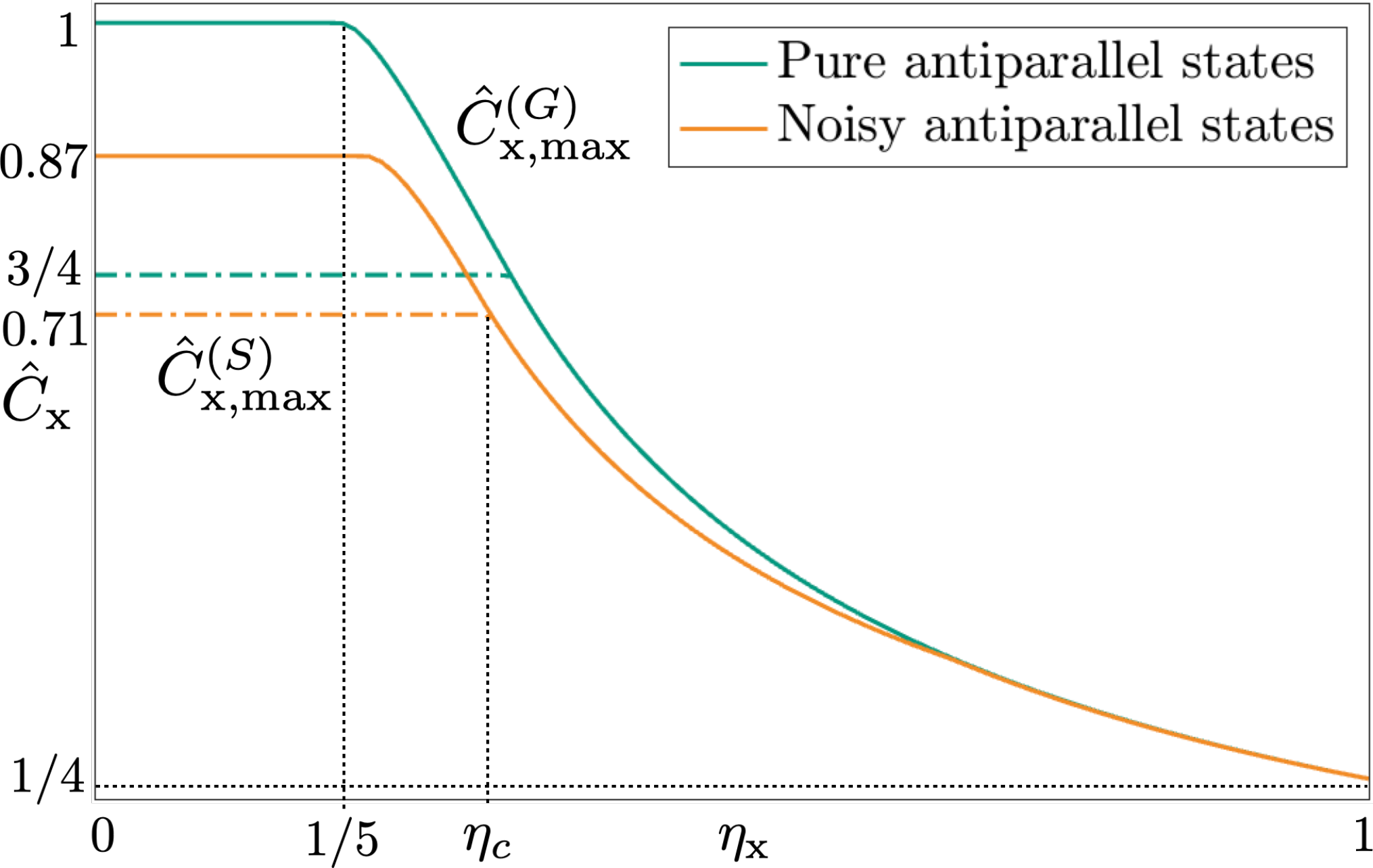}
    \caption{The certifiable maximum confidence is shown for noiseless (green) and noisy (yellow) antiparallel states. In the latter case, we choose $p=0.9$ and $\eta_c$ is slightly smaller than $1/3$.  }
    \label{fig:appfig2}
\end{figure}
Let us write $M_\x$ for the case of $\lambda=\lambda_2$ as
\bean
M_\x=\ketbra{\nu_1(\lambda_2)}+c_{1} \ketbra{\phi_\x}^{\otimes 2}+c_2 \ket{\phi_\x}\bra{\phi_\x^\perp}^{\otimes 2}
+c_2^* \ket{\phi_\x^\perp} \bra{\phi_\x}^{\otimes 2}+c_3 \ketbra{\phi_\x^\perp}^{\otimes 2}
\eean
where $\{c_i\}_{i=1}^3$ are complex numbers and
\bean
\ket{\nu_1(\lambda_2)}=\cos\xi\ket{\phi_\x}\otimes \ket{\phi_\x^\perp}+\sin\xi\ket{\phi_\x^\perp}\otimes \ket{\phi_\x}
\eean
with
\bean
\xi=\tan^{-1} \left ( \frac{\lambda_2 p}{3+\sqrt{\lambda_2^2 p^2+9}} \right ).
\eean
The calculation shows that $M_\x^{\Gamma}\geq 0$, where $\Gamma$ denotes partial transpose, if and only if 
\bea
c_1c_3 \geq \cos^2\xi \sin^2\xi \geq |c_2|^2. \label{eq:appppt}
\eea
We want to find the critical outcome rate $\eta_c$ such that the measurement must be entangled whenever $\eta_\x<\eta_c$. Note that the outcome rate by $M_\x$ is
\bean
\eta_\x&=&\tr[\rho M_\x]\\
&=&\frac{1}{12}(3+p^2- p^2 \cos\xi \sin\xi+(3-p^2)(c_1+c_3))
\eean
and when $c_1=c_3=\cos\xi\sin\xi$, we have
\bean
\eta_\x=\eta_c.
\eean
To show that $M_\x$ can be separable for $\eta_\x \geq \eta_c$, we consider an anstaz for the parameters $\{c_i\}$ such that $\cos\xi\sin\xi \leq c_1=c_3\leq1$ and $c_2=0$. Explicitly, for any $\eta_\x \geq \eta_c$, the parameters are
\bea
c_1=c_3=\frac{12\eta_\x -3-p^2(1-\cos\xi\sin\xi)}{6(1-p^2)}. \nonumber
\eea
They satisfy the inequalities in Eq. \eqref{eq:appppt}, and therefore $M_\x$ is separable. 

Finally, let us show that $M_\x$ must be entangled for $\eta_\x < \eta_c$. 
That is, we need to find the minimum value of $\eta_\x$ for which there exists a separable operator solution to $M_\x$. Since $\eta_\x$ 
is independent of $c_2$, the optimization to find the minimum $\eta_\x$ is given as
\bean
&&\mathrm{minimize~} c_1+c_3\\
&&\mathrm{subject~to~} c_1c_3\geq \cos^2\xi \sin^2\xi, 0 \leq c_1,c_3,\leq 1.
\eean
As the optimal parameters occur at the boundary, we replace $c_3=\frac{\cos^2\xi \sin^2\xi}{c_1}$. Solving $\frac{d }{dc_1}(c_1+\frac{\cos^2\xi \sin^2\xi}{c_1})=0$, we find
\bean
c_1=c_3=\cos\xi\sin\xi
\eean
Therefore, when $\eta_\x<\eta_c$, $M_\x$ must be entangled. 

\subsection*{VI. NLWE in inconclusive outcomes for noisy antiparallel states}
We have shown so far NLWE from the noisy antiparallel states can be certified from certifiable confidence and outcome rates. In this Appendix, we show that NLWE may also arise from inconclusive outcomes and can be certified even when certification from conclusive outcomes fails.
We consider noisy antiparallel states,
\bea
\rho_{\x}=\left(p|\phi_{\x}\rangle \langle \phi_{\x}| + \frac{(1-p)}{2}\I \right) \otimes \left(p|\phi_{\x}^{\perp}\rangle \langle \phi_{\x}^{\perp}| + \frac{(1-p)}{2}\I \right). \nonumber
\eea
Recall that the global MCM is given by $M_\x^{(G)}=a_\x \ketbra{\varphi_\x}$ in Eq. \eqref{app:globalmcm}, where
\bea 
\ket{\varphi_\x}= \cos\theta \ket{\phi_\x}\otimes \ket{\phi_\x^\perp}+\sin\theta\ket{\phi_\x^\perp}\otimes \ket{\phi_\x}. \nonumber
\eea
and $\theta$ is determined by $p$ as in Eq. \eqref{eq:apptheta}. We consider a noisy POVM element,
\bea
\widetilde{M}_\x^{(G)}=a (\ketbra{\varphi_\x}+\gamma \id) \nonumber
\eea
where $\gamma \geq 0$ decreases the value of confidence. If $\gamma$ is large enough, such that $\frac{1}{4}\frac{\tr[\rho_\x \widetilde{M}_\x^{(G)}]}{\tr[\rho \widetilde{M}_\x^{(G)}]}\leq C_{\x,\max}^{(S)}$, then NLWE cannot be certified from outcome $\x$, the condition which can be explicitly written as 
\bea
\gamma \geq \frac{1}{4}(1+p^2+ 2p \cos 2 \theta ) (1-\frac{C_{\x,\max}^{(S)}}{C_{\x,\max}^{(G)}})(4C_{\x,\max}^{(S)}-1)^{-1}.  \nonumber
\eea

The inconclusive rate by this POVM is
\bea
\eta_0=1-a(1+4 \gamma + \frac{p^2}{3}(1-2 \sin 2 \theta)). \nonumber
\eea
To find the minimum probability of inconclusive outcomes by this measurement, we maximize $a$ such that $\{M_\x^{(G)}\}$  with an additional POVM element forms a POVM. The optimization is formulated as follows,
\bea
&&\text{maximize } a\nonumber\\
&&\mathrm{subject~to~} \sum_{\x=1}^4 \widetilde{M}_\x^{(G)} \leq \id \nonumber. 
\eea
The inequality constraint can be rewritten as 
\bea
\frac{a}{1-4 a \gamma} \Pi\leq \id, \nonumber
\eea
where $\Pi=\sum_{\x=1}^4\ketbra{\varphi_\x}$. Denote $\nu_i$ as the eigenvalues of $\Pi$, 
which has two distinct eigenvalues 
\bea
\nu_1=\frac{2}{3}(1+\sin 2\theta),~\nu_2=2(1-\sin 2 \theta). \nonumber
 \eea
The inequality constraint reduces to
\bea
a \leq \frac{1}{4 \gamma + \nu_i}, i=1,2.\nonumber
\eea
Therefore, the maximum $a$ that satisfies this inequality is
\bea
a&=&\frac{1}{4 \gamma + \max\{\nu_i\}} \nonumber\\
&=&\begin{cases}
3( 12 \gamma+2+2\sin 2\theta)^{-1}, \text{ if } \theta \geq \frac{\pi}{12}\\
(4 \gamma + 2 - 2 \sin 2 \theta)^{-1}, \text{ if } \theta < \frac{\pi}{12} \nonumber
\end{cases}
\eea
We remark that when $\theta=\pi/12$, the noisy POVM elements $\{\widetilde{M}_\x^{(G)}\}$ forms a POVM, i.e. $\eta_0^{(G)}=0$.

Let us now consider a separable MCM given in Eq. \eqref{eq:appsepmcm},
\bea
M_\x^{(S)}=b_\x \ketbra{\Psi_\x} \nonumber
\eea
where $\ket{\Psi_\x}=\ket{\phi_\x} \otimes \ket{\phi_\x^\perp}$.
To find the minimum value of inconclusive rate, which is given by
\bea
\eta_0=1-\sum_{\x=1}^4 \tr[\rho M_\x^{(S)}]=1-\frac{1}{12}(3+p^2)\sum_{\x=1}^4 b_\x, \nonumber
\eea
we formulate the following SDP,
\bea
\text{maximize } &&\sum_{\x=1}^4 b_\x \nonumber\\
\text{subject to } &&\sum_{\x=1}^4 b_\x \ketbra{\Psi_\x} \leq \id, \nonumber\\
&& b_\x \geq 0,~\forall \x.\nonumber
\eea
We take an approach of the complementary problem, which is to find the parameters that satisfy the optimality conditions. 
To derive the optimality conditions, we first construct the Lagrangian with the dual parameters $\mu_\x, K \geq 0$,
\bean
\mathcal{L}&=&\sum_{\x=1}^4 b_\x + \tr[K(\id-\sum_{\x=1]}^4 b_\x \ketbra{\Psi_\x})]+\sum_{\x=1}^4 b_\x \mu_\x\\
&=&\sum_{\x=1}^4 b_\x (1-\tr[K \ketbra{\Psi_\x}] + \mu_\x) + \tr[K].
\eean
From this, the optimality conditions can be derived as follows,
\bea
&&1-\tr[K \ketbra{\Psi_\x}] + \mu_\x=0, \nonumber\\
&&\tr[K M_0]=0, \text{ and }\mu_\x b_\x =0, \forall \x, \nonumber
\eea
where $M_0=\id-\sum_{\x=1}^4 b_\x \ketbra{\Psi_\x}$. 
We employ an ansatz, $b_\x=b>0$ for all $\x$. Note that $M_0$ cannot have full rank. Then, by analyzing of the eigenvalues of $M_0$, we see that $b=\frac{1}{2}$ makes $M_0$ a rank-three operator,
\bea
M_0=\frac{2}{3} \Pi_{\mathrm{sym}} \nonumber
\eea
where $\Pi_{\mathrm{sym}}$ denotes a projection onto the symmetric subspace. Therefore $K$ must take the form
\bea
K=\gamma \ketbra{\psi^{-}} \nonumber,
\eea 
where $\ket{\psi^-}=\frac{1}{\sqrt{2}} (\ket{01}-\ket{10})$ and $\gamma >0$. We choose $\mu_\x=0, \forall \x$ and $\gamma=1/|\langle \psi^- | \Psi_\x \rangle |^2=2$. Since these parameters satisfy all of the optimality conditions, we conclude that
\bea
b_\x=\frac{1}{2}, \forall \x \nonumber
\eea
are the optimal parameters. 
With this, the minimum inconclusive rate by the separable MCM is obtained as
\bea
\eta_{0, \min} ^{(S)}=\frac{1}{2}-\frac{p^2}{6} \nonumber.
\eea

\end{document}